\documentclass[showkeys,eqsecnum,twocolumn,showpacs,preprintnumbers,amssymb,aps]{revtex4}
\usepackage{graphicx}
\usepackage{slashbox}
\usepackage{dcolumn}
\usepackage{bm}
\usepackage{latexsym,epsfig}

\begin{document}

\title{The Search For A Permanent Electric Dipole Moment Using Atomic Indium}

\author{B. K. Sahoo \footnote{bijaya@prl.res.in}}

\affiliation{Theoretical Physics Division, Physical Research Laboratory, Ahmedabad-380009, India}

\author{R. Pandey and B. P. Das}

\affiliation{Theoretical Astrophysics Group, Indian Institute of Astrophysics, Bangalore-560034, India}

\date{Received date; Accepted date}
\vskip1.0cm

\begin{abstract}
We propose indium (In) as a possible candidate for observing the 
permanent electric dipole moment (EDM) arising from the violations of parity
(P) and time-reversal (T) symmetries. This atom has been laser cooled and
therefore the measurement of its EDM has the potential of improving on the
current best EDM limit for a paramagnetic atom which comes from thallium. We
report the results of our calculations of the EDM enhancement factor
due to the electron EDM and the ratio of the atomic EDM to the electron-nucleus scalar-pseudoscalar
(S-PS) interaction coupling constant in In in the framework of the relativistic coupled cluster
theory. It might be possible to get new limits for the electron EDM and the 
S-PS CP violating coupling constant by combining the results of our calculations
 with the measured value of the EDM of In when
it is available. These limits could have important implications for the
standard model (SM) of particle physics. 
\end{abstract}

\maketitle

It is now widely recognized that atomic electric dipole moments (EDMs) arising 
from the violations of parity (P) and time-reversal (T) symmetries can provide 
important information about new physics beyond the standard model (SM) 
\cite{barr, ginges}. T violation implies CP violation via the CPT theorem 
\cite{luders}. The dominant sources of the EDM of a paramagnetic atom are the 
EDM of an electron and the scalar-pseudoscalar (S-PS) interaction between the 
electron and the nucleus which violates P as well as T symmetries 
\cite{sandars}. Atomic EDMs due to the electron EDM and the S-PS electron-nucleus 
interaction can shed light on CP violation in the leptonic and semi-leptonic sectors
\cite{barr,pilaftsis}. The origin of both of these kinds of CP violations are not well 
understood. 

The best limit on the EDM of a paramagnetic atom currently comes from thallium (Tl) 
\cite{regan,dzuba,sahoo,nataraj}.
A new generation of EDM experiments on the alkali atoms like rubidium \cite{weiss},
caesium \cite{weiss,bijlsma,amini} and francium \cite{sakemi} based on the techniques
of laser cooling and trapping are currently underway. These experiments in principle have 
the advantages of both the beam and the cell experiments \cite{romalis}. In particular it is 
possible
to apply large electric fields and the coherence times are long in these experiments 
\cite{romalis}. The projected precision of the 
current alkali EDM experiments are about two orders of magnitude better than that of the Tl
experiment \cite{weiss,bijlsma,amini}. Indium (In), which is homologous to Tl appears to be an 
attractive candidate for the search of a permanent EDM for essentially two reasons: 
First, this atom has been laser cooled \cite{kloeter}
 and second, the EDM enhancement factor due to the electron EDM and the ratio of the atomic EDM to the S-PS coupling constant can be calculated more accurately than that
of Tl due to its relatively smaller size. To demonstrate the second point more elaborately,
we have carried out {\it ab initio} calculations for the corresponding factors
due to the EDM of
electron and the S-PS electron-nucleus interaction in In and discuss the role
of different correlation effects in these properties here.

Following the work of Sandars \cite{sandars} and its extension
\cite{das,mukherjee,nataraj1}, the effective atomic EDM Hamiltonian due to the electron 
EDM can be written as
\begin{eqnarray}
H_{EDM}^{e}&=&2icd_{e}\sum_{j}\beta_{j}\gamma_{j}^{5}p_{j}^{ 2}
\label{eqn1}
\end{eqnarray}
and the S-PS interaction Hamiltonian is given by
\begin{eqnarray}
H_{EDM}^{S-PS}&=&\frac{iG_{F}}{\sqrt{2}}C_{S}\sum_{j}\beta_{j}\gamma_{j}^{5}\rho_{n}(r_{j}),
\label{eqn2}
\end{eqnarray}
where $d_{e}$ is the intrinsic $e$-EDM,
$\gamma^{5}$ is a pseudo-scalar Dirac matrix, $C_{S}$ is the dimensionless 
S-PS constant and $\rho_{n}(r_{j})$ is the $j$th electron density over
the nucleus. 

The above interaction Hamiltonians mix atomic states of opposite parities 
but with the same angular momentum. They can be treated as first order 
perturbations as their strengths are sufficiently weak. Therefore any atomic state
with a valence electron 'v' can be expressed after the inclusion of these interactions
as
\begin{equation}
|\Psi_v \rangle = |\Psi_v^{(0)} \rangle + \lambda |\Psi_v^{(1)} \rangle ,
\label{eqn3}
\end{equation}
\noindent
where $|\Psi_v \rangle$ is the modified wave function to the original 
wave function $|\Psi_v^{(0)} \rangle$ by the first order correction
$|\Psi_v^{(1)} \rangle$. $\lambda$ represents the weak coupling
parameter $d_e$ for $H_{EDM}^{e}$ or $C_S$ for $H_{EDM}^{S-PS}$.

The enhancement factor due to the electron EDM and the ratio of the atomic 
EDM to the S-PS coupling constant which is denoted by ${\cal R} = \frac{D_a}{\lambda}$ here for the atomic EDM $D_a$ of a state $|\Psi_v \rangle$ is given by
\begin{eqnarray}
{\cal R} &=& \frac {\langle \Psi_v^{(0)}| D |\Psi_v^{(1)} \rangle + \langle \Psi_v^{(1)}| D |\Psi_v^{(0)} \rangle } {\langle \Psi_v^{(0)}|\Psi_v^{(0)} \rangle } ,
\label{eqn4}
\end{eqnarray}
where D is the electric dipole (E1) operator. With the explicit form of
$|\Psi_v^{(1)} \rangle$, we get
\begin{eqnarray}
{\cal R} &=& \sum_{I \ne v} \frac {\langle \Psi_v^{(0)}| D |\Psi_I^{(0)} \rangle \langle \Psi_I^{(0)}| H_{EDM} |\Psi_v^{(0)} \rangle } {(E_v - E_I) \langle \Psi_v^{(0)}|\Psi_v^{(0)} \rangle} \nonumber \\
&&+ \sum_{I \ne v} \frac {\langle \Psi_v^{(0)}| H_{EDM} |\Psi_I^{(0)} \rangle \langle \Psi_I^{(0)}| D |\Psi_v^{(0)} \rangle } {(E_v - E_I) \langle \Psi_v^{(0)}|\Psi_v^{(0)} \rangle} ,
\label{eqn5}
\end{eqnarray}
where $I$ represents the intermediate states and $H_{EDM}$ is one of the EDM
interactions given above.

The above expression depends explicitly on 
E1 matrix elements, excitation energies (EEs) and matrix elements of $H_{EDM}$.
However, it is possible to consider only a finite number of excited states in the evaluation
of this quantity if a sum-over-states approach is used. In order to circumvent this problem, we solve
the first order perturbed wave function in a similar approach of
\begin{eqnarray}
(H^{(0)} - E_v^{(0)})|\Psi_v^{(1)}\rangle = - H_{EDM} |\Psi_v^{(0)}\rangle .
\end{eqnarray}
In the above expression, $H^{(0)}$ is the atomic Hamiltonian and $E_v^{(0)}$ is the energy for the
state $|\Psi_v^{(0)}\rangle$. It is possible to estimate the accuracy of the
corresponding ${\cal R}$ values by calculating the properties required
to determine Eq. (\ref{eqn5}) for the dominant intermediate states.

We employ the coupled-cluster theory in the relativistic framework (RCC theory)
 to evaluate $\vert \Psi_v^{(0)} \rangle$ and $\vert \Psi_v^{(1)} \rangle$
as described in \cite{sahoo,nataraj,mukherjee,nataraj1}. These wave functions can
be expressed as
\begin{eqnarray}
\vert \Psi_v^{(0)} \rangle &=& e^{T^{(0)}}\{1 + S_v^{(0)} \} \vert\Phi_v\rangle \ \ \
\label{eqn6} \\
\text{and} \ \ \
\vert \Psi_v^{(1)} \rangle &=&  e^{T^{(0)}}\{ T^{(1)} \left (1 + S_v^{(0)} \right ) + S_v^{(1)} \} \vert\Phi_v\rangle \ \ \ \
\label{eqn7}
\end{eqnarray}
where $\vert\Phi_v\rangle$ is the Dirac-Fock (DF) wave function obtained by appending the valence
electron $v$ to the closed-shell ($[4d^{10}]\,5s^2$ in the present case) 
reference wave function, $T^{(0)}$ and $S_v^{(0)}$ are the excitation operators 
for the core and valence electrons in the unperturbed case, where as, $T^{(1)}$ and 
$S_v^{(1)}$ are their first order corrections. The atomic wave functions are 
calculated using the Dirac-Coulomb (DC) Hamiltonian given by,
\begin{eqnarray}
H_0 &=& \sum_i \{ c {\bf \alpha}_i \cdot {\bf  p}_i + (\beta_i - 1)m_i c^2 + V_{n}(r_i)\} \nonumber \\ & &  + \sum_{i<j} V_C(r_{ij}), 
\label{H0}
\end{eqnarray}
where ${\bf \alpha}$ and $\beta$ are Dirac matrices, ${\bf p}$ is the momentum operator, 
$V_{n}(r)$ is the nuclear potential and $V_C(r)$ is the Coulomb potential.

We consider only single and double excitations in the 
expansion of the RCC wave functions (CCSD approximation), by defining,
\begin{eqnarray}
T = T_1 + T_2 \hspace{0.5cm} \text{and} \hspace{0.5cm} S_v = S_{1v} + S_{2v},
\label{eqn8}
\end{eqnarray}
for both the perturbed and unperturbed operators. Further, we construct triple
excitation operators for $S_v^{(0)}$ as,
\begin{eqnarray}\label{s30}
S_{vab}^{pqr,(0)} &=&  \frac{\widehat{H_0\, T_2^{(0)}} + \widehat{H_0\, S_{2v}^{(0)}}}{\epsilon_v + \epsilon_a + \epsilon_b - \epsilon_p - \epsilon_q - \epsilon_r},
\end{eqnarray}
which are used to evaluate the CCSD amplitudes iteratively. This is referred to 
as CCSD(T) approximation. Here, $\epsilon_i$ is the single particle energy of 
an orbital $i$.

Hence, ${\cal R}$ in the RCC theory is given by
\begin{eqnarray}\label{enfact}
{\cal R} &=& \frac{\langle \Phi_v |\{1+ S_v^{(0)^\dagger}\} \overline{D^{(0)}} \{ T^{(1)} (1+ S_v^{(0)}) + S_v^{(1)}\} |\Phi_v \rangle }{\langle \Phi_v |\, e^{T^{(0)^\dagger}}\, e^{T^{(0)}} + S_v^{(0)^{\dagger}}\, e^{T^{(0)^{\dagger}}}\, e^{T^{(0)}}S_v^{(0)}\, | \Phi_v \rangle}  \nonumber \\ 
&& + \ cc
\end{eqnarray}
where the dressed operator $\overline{D^{(0)}} = e^{{T^{(0)}}^\dagger}\, D\, e^{T^{(0)}}$,
 $D = e\, {\bf r}$ is the E1
operator due to the applied electric field and $cc$ represents the complex
conjugate terms. The procedure for the calculation of the above expression
is discussed elsewhere \cite{sahoo,nataraj,mukherjee,nataraj1}.

\begin{table}[h]
\caption{Enhancement factor due to electron EDM and the ratio of the atomic EDM to the S-PS coupling constant denoted by ${\cal R}$ to the ground state of In due to $d_e$ (dimensionless) and S-PS (in $\frac{G_{F}}{\sqrt{2}}C_{S}/A$; $A$ is the atomic mass number) interactions obtained using DF and RCC methods.}
\begin{center}
\begin{tabular}{lccc}\hline\hline
Source & DF & CCSD & CCSD(T) \\ \hline 
 & & \\
${d_e}$ & $-49.53$ & $-82.35$ & $-82.37$ \\
S-PS & $-31.10$ & $-52.59$ & $-52.60$ \\ \hline\hline
\end{tabular}
\end{center}
\label{tab1}
\end{table}
The single particle orbitals in our calculations are a linear combination of Gaussian type 
of orbitals (GTOs). They are optimized by 
comparing the energies and the radial integrals of these orbitals with those
obtained numerically from GRASP2 \cite{grasp}. We have allowed
excitations from all the occupied orbitals to unoccupied bound and continuum orbitals
with a maximum energy of 1500au. This space is sufficiently large for the convergence 
of the results of our calculations. Orbitals up to $l=4$ were included in the active space 
after
observing that the contributions from the orbitals with higher angular momenta were very small;
inclusion of these orbitals would have been computationally expensive with little or no effect
on the overall results.

\begin{table}[t]
\caption{Contributions to ${\cal R}$ (with same unit as in Table \ref{tab1}) 
from different CCSD(T) terms. $\overline{D_{o.b.}^{(0)}}$ represents
effective one-body terms from $\overline{D^{(0)}}$, terms containing bare $D$ 
operator are the effective two-body terms from $\overline{D^{(0)}}$, "Others" 
and "Norm" give contributions from other non-linear terms and normalization 
of the wave function, respectively.}
\begin{center}
\begin{tabular}{lcc}\hline\hline
Term & from $d_e$ & from S-PS \\ \hline 
 & & \\
\multicolumn{3}{c}{From DF}  \\
$(D H_{EDM})_c+cc$ & $-17.56$ & $-11.03$ \\
$(D H_{EDM})_v+cc$ & $-31.98$ & $-20.07$ \\ \hline
 & & \\
\multicolumn{3}{c}{Important RCC terms} \\
$\overline{D_{o.b.}^{(0)}}T_{1}^{(1)}+cc$ & $-39.79$ & $-25.33$ \\
$\overline{D_{o.b.}^{(0)}}S_{1v}^{(1)}+cc$ & $-44.59$ & $-28.24$ \\
$\overline{D_{o.b.}^{(0)}}S_{2v}^{(1)}+cc$ & 10.10 & $6.39$ \\
 & & \\
\multicolumn{3}{c}{Nonlinear RCC terms} \\
${T_{1}^{(1)}}^{\dagger}\overline{D_{o.b.}^{(0)}}S_{2v}^{(0)}$ & $-4.12$ & $-2.62$ \\
${S_{1}^{(1)}}^{\dagger}\overline{D_{o.b.}^{(0)}}S_{1}^{(0)}+cc$ & 3.55 & $2.27$ \\
${S_{2}^{(0)}}^{\dagger}\overline{D_{o.b.}^{(0)}}S_{1v}^{(1)}+cc$ & $-2.06$ & $-1.29$\\
${T_{1}^{(1)}}^{\dagger}DT_{2}^{(0)}+cc$ & $-4.74$ & $-3.02$ \\
${T_{2}^{(0)}}^{\dagger}DT_{1}^{(1)}S_{2}^{(0)}$ & $-1.24$ & $-0.78$ \\
${T_{2}^{(0)}}^{\dagger}DS_{2}^{(1)}+cc$ & $0.86$ & $0.56$ \\
${S_{2v}^{(0)}}^{\dagger}DS_{2v}^{(1)} + cc $ & $-0.97$ & $-0.61$ \\
Others & 0.21 & $0.17$\\
Norm & 0.83 & $0.52$ \\ \hline\hline
\end{tabular}
\end{center}
\label{tab2}
\end{table}

In Table \ref{tab1}, we present the ${\cal R}$ values for both $d_e$ and S-PS
interactions at the DF, CCSD and CCSD(T) levels. There is a significant
difference between the DF and CCSD(T) results highlighting the importance
of strong correlation effects for the reported ${\cal R}$ factors of this system. Small difference
between the CCSD and CCSD(T) results suggests that the contributions from
the triple excitations are small. We give below contributions from various
CCSD(T) terms to understand the roles of different correlation effects
in this property.

\begin{figure}[t]
\includegraphics[width=8.5cm]{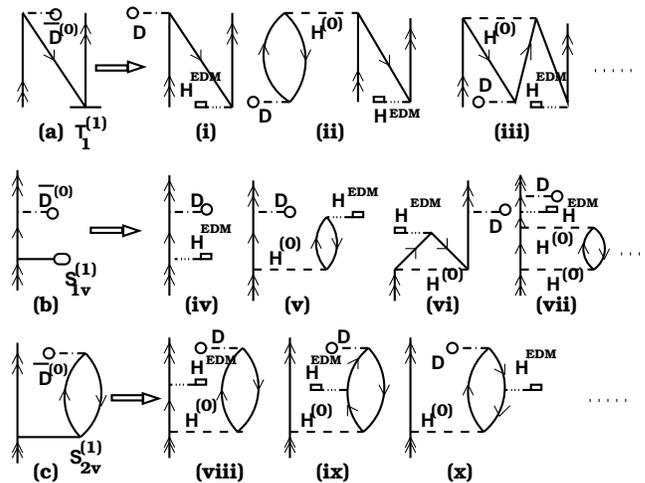}
\caption{Break down of important perturbed CCSD(T) diagrams into lower order
many-body perturbation diagrams. Lines with single arrow showing up and
down wards represent the occupied and unoccupied orbitals, respectively.
Line with a double arrow represent the valence orbital.}
\label{fig1}
\end{figure}

In Table \ref{tab2}, we present the contributions from the core and the virtual
orbitals to the factors ${\cal R}$ at the DF level and individual
contributions from different CCSD(T) terms. The CCSD(T) contributions in this 
table have been classified as important terms referring to terms whose 
contributions are large (Goldstone diagrams for these terms are shown in
Fig. \ref{fig1}) and nonlinear terms whose contributions are relatively small.
Fig. \ref{fig1}(a) representing $\overline{D^{(0)}}T_1^{(1)}$ involves the
lowest order DF contributions due to the core orbitals (Fig. \ref{fig1}(i))
and some of the higher order core-polarization correlation diagrams 
(Fig. \ref{fig1}(ii,iii) and more). By comparing the DF contributions from 
core orbitals and $\overline{D^{(0)}}T_1^{(1)}$, it is obvious that these 
core-polarization correlation effects contribute significantly; larger
than the DF core contribution. The largest contributions come from the
$\overline{D^{(0)}}S_{1v}^{(1)}$ term and as shown in Fig. (\ref{fig1}(b)),
it contains the lowest order DF contribution due to the virtual orbitals.
It also includes many important core-polarization (Fig. \ref{fig1}(v,vi))
and pair (Fig. \ref{fig1}(vii)) correlation diagrams. However, the net 
correlation contribution due to this diagram is not as large as it is
from the higher order diagrams in $\overline{D^{(0)}}T_1^{(1)}$. The other 
important term that has been considered is $\overline{D^{(0)}}S_{2v}^{(1)}$
and it accounts for mainly a particular class of core polarization effects.

We now consider some of the quantitative aspects of the correlation effects in the above electron EDM enhancement factor and the ratio of the atomic EDM to the S-PS coupling constant.
These effects in In are significantly larger than those in the alkalies, 
but their cancellations are not as severe as in the case of Tl. The correlation effects 
for both the above quantities are one and a half times the total DF results. The core 
correlation effects are very strong, because of the large overlap between the wave functions 
of the valence
$5p_{1/2}$ and the outermost core $5s$ orbitals and the small energy difference between them. 
At the DF level, the contributions from the virtual orbitals are larger than those of  
the core orbitals. The total contributions from both $\overline{D^{(0)}}T_1^{(1)}$ and $\overline{D^{(0)}}S_{1v}^{(1)}$ are  
comparable. The contribution from $\overline{D^{(0)}}S_{2v}^{(1)}$ is significant, but with 
opposite sign. These contributions are from the singly excited perturbed
states; mainly from the $[4d^{10}]5s5p^{2}$ state. There are also 
some significant contributions from the higher order RCC terms; especially 
${T_{1}^{(1)}}^{\dagger}DT_{2}^{(0)}$, ${S_{1}^{(1)}}^{\dagger}\overline{D^{(0)}}S_{1}^{(0)}$, 
${S_{2}^{(0)}}^{\dagger}\overline{D^{(0)}}S_{1v}^{(1)}$  and 
${T_{1}^{(1)}}^{\dagger}\overline{D^{(0)}}S_{2v}^{(0)}$ through some of the 
core-polarization and pair-correlation effects. Contributions due to the
normalization of the wave function (norm) and terms containing non-linear
terms in $T^{(0)}$ and $T^{(1)}$ are small.
\begin{table}[t]
\caption{Comparison of the excitation energies (in $cm^{-1}$), E1 matrix element and hyperfine structure constants of low-lying states in In with the available experimental results and all order SD calculations. Uncertainties from our calculations are given in the parentheses.}
\begin{center}
\begin{tabular}{cccc}
\hline \hline
           & This Work & All order SD \cite{safronova} & Experiment \\
\hline
Transition & \multicolumn{3}{c}{Excitation energies} \\ \cline{2-4} \\ 
                 &      & \\
$6\,s \rightarrow 5\,p_{1/2}$ & 24290(80) & 23747 & 24372.956 \cite{moore} \\
$7\,s \rightarrow 5\,p_{1/2}$ & 36217(90) & 35808 & 36301.84 \cite{moore} \\
$8\,s \rightarrow 5\,p_{1/2}$ & 40552(100) & 40126  & 40637.0 \cite{moore} \\
$9\,s \rightarrow 5\,p_{1/2}$ & 42640(115) & 42238 & 42719.0 \cite{moore} \\
                 &      & \\
Transition & \multicolumn{3}{c}{E1 matrix elements} \\ \cline{2-4} \\ 
                 &      & \\
$6\,s \rightarrow 5\,p_{1/2}$ & 1.91(1) & 1.91 & 1.92(8) \cite{andesen} \\
$7\,s \rightarrow 5\,p_{1/2}$ & 0.56(2) & 0.54 &  \\
$8\,s \rightarrow 5\,p_{1/2}$ & 0.31(2) & 0.09 &  \\
$9\,s \rightarrow 5\,p_{1/2}$ & 0.19(2) &   &  \\
                &      & \\
State  & \multicolumn{3}{c}{$A_{hyp}$ of $^{115}$In } \\ \cline{2-4}
                  &      & \\
$5\,p_{1/2}$  & 2256(30) & 2306  & 2282(40) \cite{eck} \\
$6\,s$  & 1611(50) & 1812  & 1687.2(6) \cite{george} \\
$7\,s$  &  516(30) & 544.5 &  541.1(3) \cite{george} \\
$8\,s$  &  234(20) & 240.8 &   \\
$9\,s$  &  106(10) & 128.1 &   \\
\hline \hline
\end{tabular}
\end{center}
\label{tab3}
\end{table}

It is possible to get a sense of the
 accuracies of the individual
quantities that appear in Eq. (\ref{eqn5}) by comparing their calculated values  
with their corresponding experimental results. Experimental values for the
EEs are available up to very high accuracy \cite{moore}
and experimental values of the E1 matrix elements can be extracted from the 
lifetime measurements of the available $s$ states. 
The matrix elements of $H_{EDM}$ cannot be measured directly. However, 
the accuracy of this quantity  can be indirectly estimated
from the square root of the product of the magnetic dipole hyperfine
structure constants ($A_{hyp}$) of the appropriate states \cite{sahoo,nataraj}. 
 
We present the results for the EEs, E1 matrix elements and the hyperfine structure constants 
of the ground and excited
$s$ states in Table \ref{tab3} that contribute significantly to ${\cal R}$.
Experimental values for the EEs are available for most of the excited states, 
but the hyperfine structure constants are experimentally known only for the
$5p_{1/2}$, $6s$ and $7s$ states. We obtain the E1 matrix element of the
$6s \rightarrow 5p_{1/2}$ transition as $1.92(8)au$ by combining the 
experimental values of the lifetime of the $6s$ state ($7.5(7)s$) \cite{andesen}
 and considering the branching ratio as 2:3 of the $6s \rightarrow 5p_{1/2}$ 
and $6s \rightarrow 5p_{3/2}$ transitions \cite{kim} (Ref. \cite{bijaya}
can be referred for detailed discussion of these results). For EEs, the largest uncertainty comes from the Breit interaction followed
by the neglected basis orbitals and triple excitations. However, most of the
uncertainties to the E1 matrix elements and $A_{hyp}$s come from the latter.
All our results are in good agreement with the experimental results.
Our calculated results have been compared with those obtained by the all order
SD method \cite{safronova}, the RCC method containing only the linear terms of 
our CCSD approach, in the above table. Our EEs results are in better agreement
with accurate experimental data than those using the all order SD method for
all the excited states relevant for our present work. The corresponding
DF results are given elsewhere (see Ref. \cite{bijaya}) and it is found in
\cite{safronova} that the Breit interaction contributes very little to this 
property in this atom. Therefore, the discrepancies between the results
reported in \cite{safronova} and the present work could be due to the 
non-linear terms of the CCSD(T) method. Our $A_{hyp}$ for the $6s$ state, the
most important excited state in the calculations of ${\cal R}$, is also more 
accurate than the all order SD result. Our results for the other
quantities agree reasonably well with available experimental data.
The uncertainties in various quantities are estimated by considering the
differences between the results of the CCSD(T) and CCSD methods as the
upper limits to the contributions due to the triple excitations,
neglected relativistic effects; particularly the Breit interaction and omitted higher angular momentum
symmetry orbitals in the present calculations. After considering all
possible uncertainties, the enhancement factor due to the electron EDM and the ratio of the atomic EDM to the S-PS coupling constant
for In are estimated to be $-82(5)$ and $-53(3)$, respectively. These results are almost five and 
one and a half times smaller than those in Tl \cite{nataraj} and Cs 
\cite{nataraj1} respectively, but three times larger than those in Rb 
\cite{nataraj1}. Their accuracies can be further improved by  using the 
general RCC theory \cite{kallay}.

In conclusion, we propose In as a suitable candidate for the search of
a permanent EDM. Our theoretical studies show that accurate
calculations of the electron EDM enhancement factor and the ratio of the atomic EDM to the S-PS coupling constant  of this atom are possible.
The limits for the electron EDM and the S-PS coupling constants
that can be extracted by combining these factors with the measured value of the
EDM of this atom when it is available, could provide important information about the 
validity of the SM of particle physics.

We thank H. S. Nataraj for useful discussions. These calculations were carried
out using the HPC 3TFLOP cluster at PRL and the CDAC ParamPadma TeraFlop 
supercomputer, Bangalore.

\end{document}